# The Linear and Non-linear Magnetic Response of a Tri-Uranium Single Molecule Magnet

B.S. Shivaram[1], Eric Colineau[2], Jean-Christophe Griveau[2], P.Kumar[3] and V. Celli[1]

[1]Department of Physics, University of Virginia, Charlottesville, VA. 22901

[2]European Commission, Joint Research Centre (JRC), Directorate G - Nuclear Safety and Security, Postfach 2340, D-76125, Karlsruhe, Germany

[3]Department of Physics, University of Florida, Gainesville, FL. 32611

12/19/2016

## ABSTRACT

We report here low temperature magnetization isotherms for the single molecule magnet, $(UO_2\text{-}L)_3$. By analyzing the low temperature magnetization in terms of $M = \chi_1 B + \chi_3 B^3$ we extract the linear susceptibility $\chi_1$ and the leading order nonlinear susceptibility $\chi_3$. We find that $\chi_1$ exhibits a peak at a temperature of $T_1 = 10.4$ K with $\chi_3$ also exhibiting a peak but at a reduced temperature $T_3 = 5$ K. At the lowest temperatures the isotherms exhibit a critical field $B_c = 11.5$ T marked by a clear point of inflection. A minimal Hamiltonian employing $S=1$ (pseudo) spins with only a single energy scale (successfully used to model the behavior of bulk f-electron metamagnets) is shown to provide a good description of the observed linear scaling between $T_1$, $T_3$ and $B_c$. We further show that a Heisenberg Hamiltonian previously employed by Carretta et al. (2013 J. Phys.Cond. Matt. **25** 486001) to model this single molecule magnet gives formulas for the angle averaged susceptibilities (in the Ising limit) very similar to those of the minimal model.

**Introduction:** The rich variety of magnetic phenomena driven by quantum mechanical effects in systems ranging from few spins to macroscopic ensembles are of interest in a variety of research areas[1]. In crystals with long range magnetic order spontaneous magnetization can arise below a critical temperature $T_c$ driven by exchange interactions. Such a phase transition is generally excluded in systems with only a few spins since they are not in the thermodynamic limit. Nevertheless single molecules can be 'magnetic' for extended periods of time even after removal of the external field. The field of single molecule magnetism (SMM) has burgeoned in the past two decades[2,3,4] with magnetic ions arranged in a variety of configurations such as triangles[5], squares[6], rings[7], and barrels[8]. Such magnetic ions in addition to their inter-ion couplings are also bonded to various ligand molecules . Dipolar interactions between the single magnetic (macro)molecules are typically small compared to the intramolecular spin interactions and magnetic measurements almost exclusively provide information about the molecule interior. To describe the measured magnetic response in such molecular magnets, typically, Heisenberg-like spin Hamiltonians are used with an appropriate number of parameters depending upon the structure and complexity of the arrangement of the magnetic ions and the possible role of the ligands. With the correct spin Hamiltonian, measurements such as magnetization, electron spin resonance (ESR) and inelastic neutron scattering (INS) spectra can be analyzed for insight into the microscopic details[9,10]. This commonly followed phenomenological approach can become impractical if the molecule is complex and many parameters have to be considered. First principles approaches such as density functional theory can be employed but they are more complex and might miss the description of strong correlation effects in unfilled d and f electron systems[11].

Interest in SMMs [12,13] is being driven by the possibility of a variety of applications such as quantum computing and information storage. SMMs containing f-electron ions are particularly interesting owing to the possibility of large anisotropy. In addition as pointed out by Carretta et al.[14] f-electron based SMMs may carry contributions from multipolar interactions in analogy to bulk actinide systems[15,16,17]. In recent work on bulk actinide metamagnets we have uncovered a number of scaling properties in their magnetic response[18,19]. Such correlations may exist in SMMs as well.

In their work Carretta *et al.* present the magnetic properties and the ESR spectra of a new f-electron based single molecule magnet, $(UO_2\text{-L})_3$, with ligand L = $C_{28}H_{21}N_4$. This is a triangular actinide molecule containing three $U^VO^+_2$ ($5f^1$) ions recently synthesized with L= 2-(4-tolyl)-1,3-bis(quinolyl)malondiiminate[20]. They interpret their results using a Heisenberg Hamiltonian with crystal field anisotropies:

$$\mathcal{H} = I(\vec{J}_1\cdot\vec{J}_2 + \vec{J}_2\cdot\vec{J}_3 + \vec{J}_3\cdot\vec{J}_1) + \sum_{i=1,3}[B_2^0\left(3J_{zi}^2 - J(J+1)\right) + B_2^2(J_{xi}^2 - J_{yi}^2)] - \vec{B}\cdot\sum_{i=1,3}g\mu_B\vec{J}_i \qquad (1)$$

Here, *I* represents the exchange interaction between the spins, $B_2^0$ is the axial anisotropy and $B_2^2$ the orthorhombic distortion in the crystal-field created by the ligands. The trimer is in the x-z plane and the $z_i$ axes are along the sides of an equilateral triangle. A peak in the linear susceptibility observed at $T_1$=10.4 K and an upturn in the magnetization isotherms apparent only at low temperatures, T < $T_1$, were both successfully modeled by this Hamiltonian. Furthermore, the model also produces the correct ESR response where experimentally only a single absorption peak was detected[14]. Since the measurements were performed on polycrystals Carretta et al. average their numerical results over several angles to fit the experimental magnetization curves. Hamiltonian (1) represents a doubly degenerate ground state and a set of three closely situated doubly degenerate first excited states for a total of 8 states. The rest of the states for three J=5/2 spins, all occupy considerably higher energies. The splitting between the ground doublet and the set of three (close in energy) excited doublet states is approximately 18 K (see Appendix). Application of a magnetic field removes the doublet degeneracy of the excited states with their resulting evolution being highly anisotropic. The lowest magnetic sublevel eventually crosses the ground state doublet at an angle-dependent critical field, leading to a sharp increase in the low-temperature magnetization, which is akin to 'metamagnetism' in bulk systems. Such a level crossing occurs at approximately the same critical field in the x-z plane, but there is no level crossing when the field is in the y-direction. Still, when the average is taken over all orientations in a powder sample, there is a marked inflection point at an effective critical field $B_c$.

**Methods:** In this paper we present further detailed measurements on the magnetic response of the $(UO_2-L)_3$ SMM. From these we demonstrate, significantly, that the scaling of the susceptibilities and the critical magnetic field established recently in bulk f-electron systems is also found here[21,22,18]. The measurements reported were performed at the Institute for Trans Uranium Elements, Karlsruhe, in fields up to 14 Tesla in a Quantum Design (QD) Physical Property Measurement System and in fields up to 7 Tesla in a QD SQUID based Magnetic Property Measurement System.

In fig.1 we present the measured isotherms of the magnetic moment, m, per molecule plotted with m/B on the y-axis and $B^2$ on the x-axis. This method of presenting the data makes it convenient to extract the linear and leading order nonlinear magnetic response given expression (2) which we find is sufficient[23] in the field range to 7 T.

$$m(T,B) = \chi_1(T)B + \chi_3(T)B^3 \qquad (2)$$

It is clear from fig. 1 that the linear susceptibility, the intercept on the y-axis, goes through a maximum, as the temperature is raised from 2K to 30 K. It is also clear that the slope of the lines which is a measure of the third order susceptibility, $\chi_3$ also goes through a maximum – it is small and positive at the lowest temperature, increases as the temperature rises, reaching a maximum and at the high temperatures is small and negative. The values for $\chi_1$ and $\chi_3$ obtained from these experimental plots are shown in fig.2 and a discussion of these results follows below.

**Results and Discussion:** The angle averaged linear susceptibility and the magnetization isotherms can be computed from (1) and indeed as shown by Carretta et al. with specific molecular parameters, I=1.05 cm$^{-1}$ and $B_2^0$= -70 cm$^{-1}$ the calculated values agree impressively with experiments  The value of $B_2^2$ in (1) is fixed from the observed ESR spectrum ($B_2^2$=56 cm$^{-1}$) and does not affect the magnetization results[14]. In order to understand <u>both</u> the linear and the nonlinear susceptibilities we adopt a much simpler approach in this paper. We start with a minimal Hamiltonian (proposed in the context of metamagnetism[18,24] in bulk f-electron systems),  $H = \Delta S_z^2 - \gamma S_z B$  where Δ and γ are adjustable parameters and $S_z$ is a component of an S=1 pseudospin. In this model the explicit expressions for the linear, third order and (see eq. A.2) fifth order susceptibilities are:

$$\chi_1 = \frac{2\gamma^2}{k_B T}\frac{1}{(2+a)} \tag{3}$$

$$\chi_3 = \frac{\gamma^4}{3k_B^3 T^3}\frac{(a-4)}{(2+a)^2} \tag{4}$$

$$\chi_5 = \frac{\gamma^6}{60 k_B^5 T^5}\frac{(a^2-26a+64)}{(2+a)^3} \tag{5}$$

where $a = e^{\Delta/k_B T}$. The merits to this minimal approach are that analytical formulae can be obtained for all the observables and this aids enormously in developing an understanding of the physical effects involved. A comparison of the results from (3) and (4) with $\Delta=16$ K and $\gamma=1.95\mu_B$ with the experimental values are given in fig.2. A surprising agreement of the minimal model with experiments is seen in fig.2 with a peak in $\chi_3$ being found at a temperature $T_3$ that is approximately half the temperature $T_1$= 10.4 K where a peak in $\chi_1$ is seen. Similarly it is easy to verify by plotting (4) and (5) that there is a peak in $\chi_5$ at a temperature roughly half of $T_3$. We now provide a rationale for these results starting from (1).

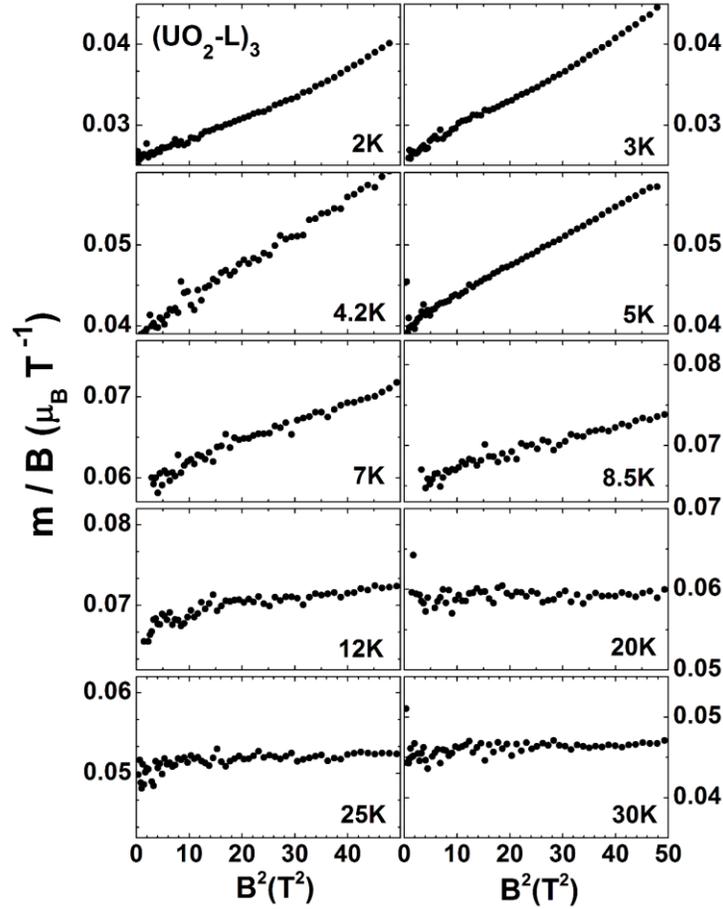

*Figure 1: The magnetic moment isotherms measured to 7 T (field sweep up) arranged to extract the linear and the third order susceptibilities in (UO₂-L)₃. Note that the linear susceptibility, the intercept in these plots, goes through a maximum at $T_1\sim 10K$. The slopes of the lines yield the third order susceptibility, $\chi_3$, which also goes through a maximum albeit at a lower temperature, $T_3=5K$.*

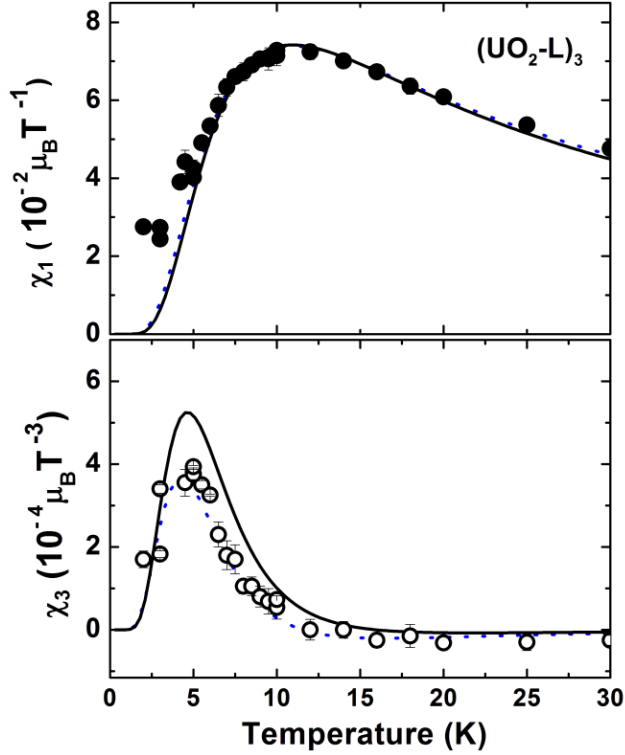

Figure 2: The linear, $\chi_1$, and the nonlinear, $\chi_3$ susceptibilities in $(UO_2\text{-L})_3$, expressed per molecule, obtained from an analysis of the plots such as in fig.1 are shown as solid and open circles respectively. The error bars shown are obtained from averaging both the field up and field down isotherms. Note the peak in $\chi_1$ at approximately 10 K and the corresponding peak in $\chi_3$ at half of this temperature. The solid black lines for $\chi_1$ and $\chi_3$ are from eqns (6) and (7) respectively. Similarly the dotted lines for $\chi_1$ and $\chi_3$ are from eqns.(3) and (4). The slight discrepancy at the lowest temperatures in $\chi_1$ is presumed due to be due to small unavoidable paramagnetic impurities.

Numerical evaluations of $\chi_1$ and $\chi_3$ for the $\mathcal{H}$ of (1) show that they depend weakly on $B_2^0$ and $B_2^2$, as long as $-B_2^0 > 70$ cm$^{-1}$ and $B_2^2 < 56$ cm$^{-1}$. We can then take the limit of a planar Ising triangle ($B_2^0 = -\infty$ and $B_2^2 = 0$) as a good approximation. Following the analysis in the Appendix we have for the averaged values of the three susceptibilities in this limit:

$$\langle \chi_1 \rangle = \frac{4\gamma^2}{k_B T} \frac{1}{(3+a)} \tag{6}$$

$$\langle \chi_3 \rangle = \frac{8\gamma^4}{5 k_B^3 T^3} \frac{(a-3)}{(3+a)^2} \tag{7}$$

$$\langle \chi_5 \rangle = \frac{\gamma^6}{35 k_B^5 T^5} \frac{(36 - 21a + a^2)}{(3+a)^3} \tag{8}$$

where $\gamma = gJ\mu_B$.

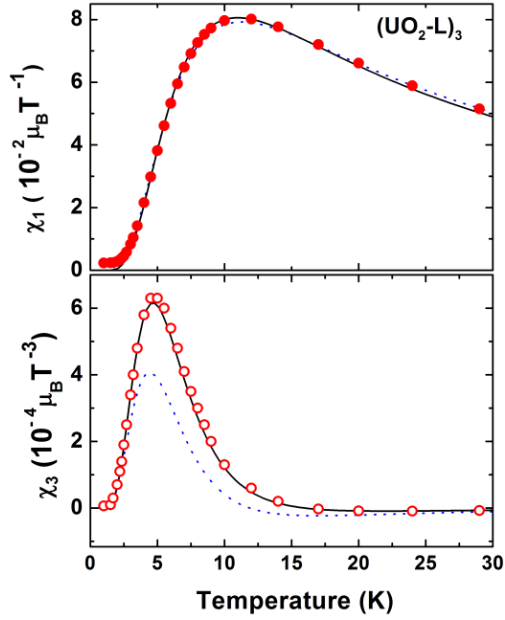

Figure 3: The values for $\chi_1$ and $\chi_3$ obtained from the full Hamiltonian (1) (solid and open circles for $\chi_1$ and $\chi_3$ respectively) compared with those obtained from the minimal model, dotted lines, and from (1) in the Ising limit, solid lines. The Ising limit parameters are $\Delta=17.5$ and $gJ=1.55$ in $\gamma = g\mu_B J$ with $J=5/2$ for the Uranium ion.

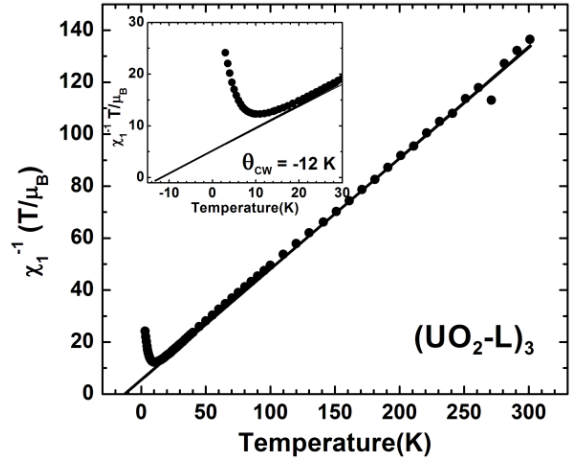

Figure 4: Shows the inverse of the measured linear susceptibility as a function of temperature. A linear fit performed on the data for $T>150$ K yields a Curie-Weiss temperature of -12 K (see inset). The minimal model on the other hand yields a $\Theta_{CW}=-5$ K compared to the Ising limit of Hamiltonian(1) which yields $\Theta_{CW}=-4$ K.

We thus see that (6), (7) and (8) bear a striking resemblance to (3), (4) and (5) with the exception of the numerical terms. We present in fig. 3 a graphical comparison of $\chi_1$ and $\chi_3$ from all three models.

Additional correlations ensuing from the above simplified approach may be examined as well. We begin with the high temperature behavior of the measured linear susceptibility. As shown in fig.4 a Curie-Weiss temperature of 12 K is obtained from an analysis of the measured high temperature $\chi_1$. By taking the T→ ∞ limit of (3) we find that the minimal model has a certain degree of "pre-built" antiferromagnetic exchange included. This degree of exchange yields a Curie-Weiss temperature, $\theta_{CW}$= $T_1/2$. We note that the parameter $\Delta$ in the Hamiltonian is related to the peak temperature $T_1$ in the model by $T_1=0.67\Delta$. From the measured value of $T_1=10.4$ K the expected $\theta_{CW}$ is 5.2 K whereas the high-T limit of (6) yields $\theta_{CW}$=-4K . Unlike a conventional magnet, the Curie Weiss temperature here should not be interpreted as due to an exchange interaction. The effect of the crystal field energy levels on the free

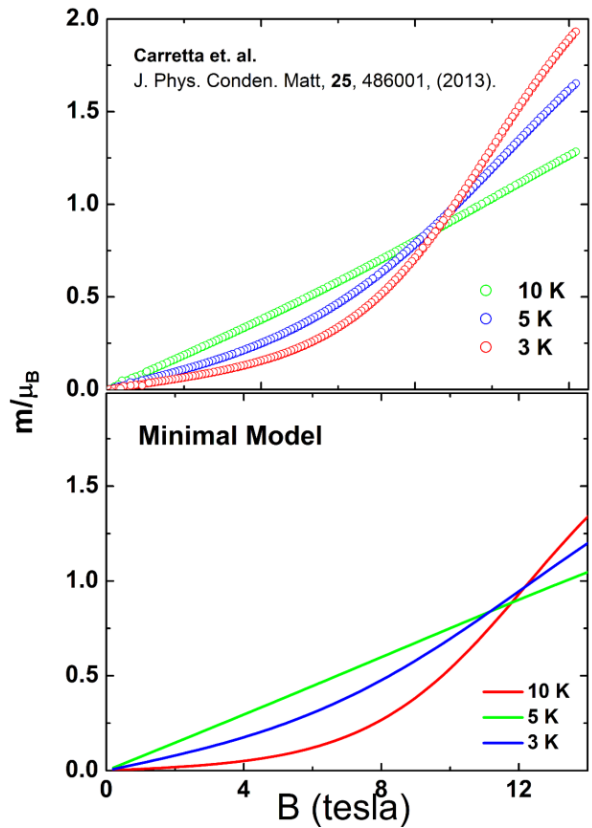

*Figure 5: The bottom panel shows the magnetization isotherms calculated from the minimal model with Δ=16 K for three temperatures as a function of the magnetic field. The single energy scale model reproduces all the salient features observed in the experiment and also modeled by Hamiltonian (1) (fig. 3 of ref.14). The data in the upper panel is from ref. 14 courtesy of Prof. S. Carretta.*

spin asymptotic high temperature susceptibility $\chi_1 \approx 1/\tau$ also plays a role.

The minimal model can also be used to evaluate the field dependent behavior of the magnetization. We show in fig.5 that the model reproduces the general behavior obtained experimentally. The magnetization isotherms show a point of inflection at low temperatures, which we label as the critical field, $B_c$. It is conveniently identified as the field (11.5 Tesla for the $UO_2$ molecule) where a maximum in the differential susceptibility occurs (inset of fig.7). This maximum is found to rise inversely with temperature in the low temperature limit, a behavior which is easily deduced from the minimal model as:

$$\frac{dm}{dB} = \frac{2[a\cosh\left(\frac{\gamma B}{k_B T}\right)+2]}{k_B T[a+2\cosh\left(\frac{\gamma B}{k_B T}\right)]^2} \tag{8}$$

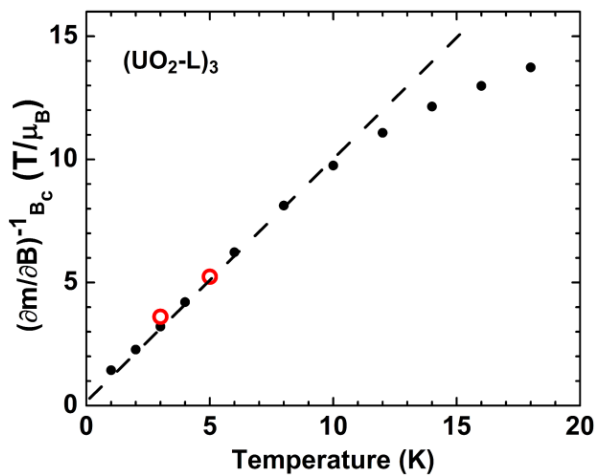

*Figure 6: Shows the inverse differential susceptibility at the critical field as a function of temperature. The dotted line is the linear behavior expected in the low temperature limit as per eq.(10). The solid black circles are from the averaged magnetization calculated from (1). The two (red) open circles are from experiments.*

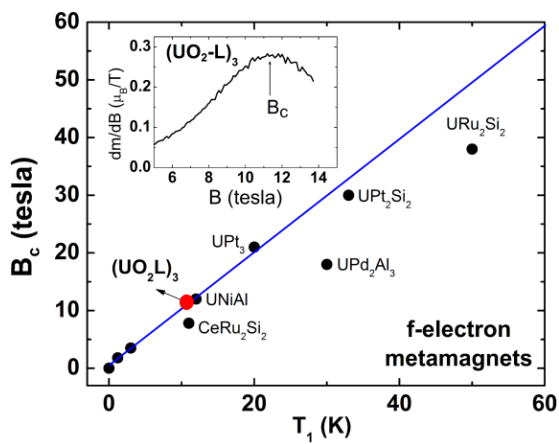

*Figure 7: Shows the critical field $B_c$ versus the temperature, $T_1$, where $\chi_1$ peaks as observed in various f-electron bulk materials. The line shown is a guide to the eye. The red dot, for the $(UO_2L)_3$ SMM, also falls on this line.*

In fig.6 we show the inverse differential susceptibility at the critical field obtained from the averaged magnetic moment vs. B curves numerically evaluated from (1). Also shown are the two experimental points available at this time - the inverse relationship implied by (8) is apparent in the figure. Such a 1/T dependence can also be derived from the expression for 'm' obtained from (1) in the

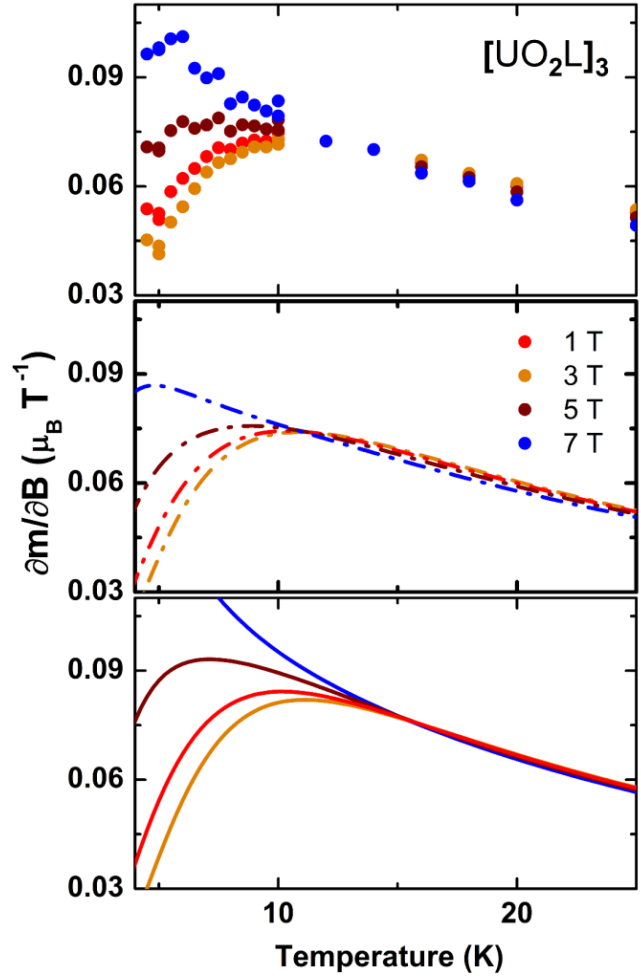

*Figure 8: The top panel shows the differential susceptibility calculated from the measured values of $\chi_1$ and $\chi_3$. The middle panel shows the differential susceptibility in the minimal model. Note the crossing of the differential susceptibility lines at different constant DC fields at a common temperature point. Such a crossing is a generic feature of strongly correlated fermion behavior. The same crossing is also seen in the Ising limit (lower panel).*

Ising limit and given in the Appendix. In fig. 7 we show the correlation of the peak temperature $T_1$ with the measured critical field for various f-electron systems. The (UO$_2$-L)$_3$ SMM falls in line with the behavior seen in the rest of the f-electron based systems (both U and Ce).

In fig. 8 we show the differential susceptibility calculated from the measured values of χ$_1$ and χ$_3$ (top panel) using the formula $\frac{\partial m}{\partial B} = \chi_1 + 3\chi_3 B^2$. Also shown in the middle panel are results from the minimal model and in the bottom panel those from the formula for m given in the Appendix (Ising approximation of (1)). The differential susceptibility curves at different fields all cross at a common temperature for fields approximately less than or equal to the critical field. It appears that as long as the temperature is not too high or the field too large the magnetization and the differential susceptibility curves cross at a common point. Such crossing points are indeed observed in several strongly correlated fermion systems and in fact it has been suggested that it is a hallmark of strong correlations[25]. It is significant that these correlations exist in a single molecule magnet as well and the apparent deviation from this crossing point either when the field is increased or when the temperature is raised is consistent with the removal of such correlations.

Single molecule magnets can incorporate many types of interaction effects. In addition to the obvious exchange interactions between the individual magnetic ions in the molecule the ligand electrons also play a significant role. Thus even in relatively simple molecules "many body" effects can exist[11]. In a large fraction of the single molecule magnets the exchange between the ions is antiferromagnetic. If this is the case such effects as magnetic frustration can also become relevant when there are an odd number of spins[26].

**Conclusion:** In conclusion in the context of the magnetic properties of a f-electron based SMM we have presented new measurements on the nonlinear susceptibilties. We have also developed a minimal single energy scale model which is able to account for all the measured susceptibilities, the linear scaling of their peak temperatures and the scaling of the critical field. Future measurements on single crystals of the (UO$_2$-L)$_3$ system as well as measurements on other d-electron based SMMs should be useful in further substantiating the model presented here. In addition to magnetic studies thermodynamic measurements on the (UO$_2$-L)$_3$ SMM such as field dependent heat capacity measurements would also be useful. The success of the minimal model should spur further understanding of SMMs particularly those with many magnetic ions with sophisticated arrangements. An explanation in simple terms as presented here should facilitate future technological developments in single molecule magnetism.

**ACKNOWLEDGEMENTS:** We are most grateful to R. Caciuffo and the Institute of Transuranium elements for enthusiastic support of this work. We acknowledge useful correspondence and conversations with S. Carretta, S. Santini, A. Amoretti and M. Mazzanti. We also acknowledge helpful conversations with Mark Meisel and Nicholas Chilton.

# Appendix

This appendix shows how the Hamiltonian, $\mathcal{H}$, eq.(1) is approximately equivalent to the minimal Hamiltonian, $\Delta S_z^2 - \gamma S_z B = 0$, for the purpose of computing the average over angles of the linear and nonlinear susceptibilities. We can show this near equivalence in the limit of very large $B_2^0$, or more precisely, $B_2^0 \gg I$, while $B_2^2/3B_2^0$ is small and can be neglected to the lowest order. Then, the uranium spins are constrained to the x-z plane and for the i$^{th}$ spin we simply have $J_{zi} = \pm\frac{5}{2}$. There are only $2^3 = 8$ states, and these are the lowest of the $6^3=216$ states of the full Hamiltonian, $\mathcal{H}$, eq.(1). The spin configurations corresponding to these states are depicted in fig. A1: the ground state doublet is on the left-most panel (top and bottom rows) and the three excited states are obtained by flipping each of the three spins, one at a time (the corresponding panels on the right). The energy difference between the ground state and any of the excited states in zero field is $\Delta_0 = 2I * \left(\frac{5}{2}\right)\left(\frac{5}{2}\right)\left\{\frac{1}{2} - \left(-\frac{1}{2}\right)\right\} = 12.5\ I$ where 1/2 and -½ are cosines of 60 and 120 degrees, respectively. With I= 1.05 cm$^{-1}$ as given by Carretta et al. $\frac{\Delta_0}{k_B} = 18.9\ K$. The correct value when $B_2^0 = -70$ cm$^{-1}$ is 18.0 K.

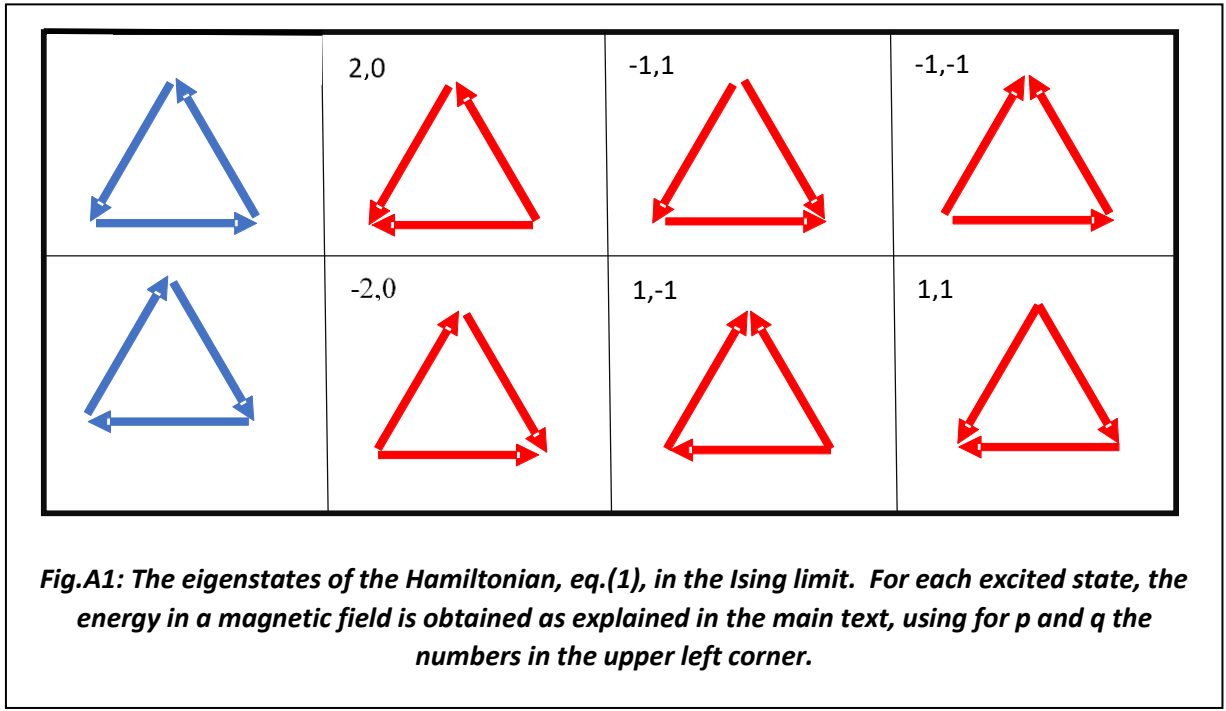

*Fig.A1: The eigenstates of the Hamiltonian, eq.(1), in the Ising limit. For each excited state, the energy in a magnetic field is obtained as explained in the main text, using for p and q the numbers in the upper left corner.*

We can conveniently set the ground state at $-\Delta_0$ and assume that the magnetic field has components $H(sin\theta cos\varphi, sin\theta sin\varphi, cos\theta)$ where $\theta$ is the polar angle perpendicular to the plane of fig. A1, and $\varphi$ is relative to the base of the triangles. Then the excited states have energies $gJ\mu_B H sin\theta (pcos\varphi + q\sqrt{3} sin\varphi)$, with p and q given in the figure. The partition function per molecule is:

$$Z = 2e^{\Delta_0/k_BT} + 2\cosh(2Bc) + 4\cosh(Bc)\cosh(\sqrt{3}Bs) \quad \text{where } c = \left(\frac{gJ\mu_B}{k_BT}\right)\sin\theta\cos\varphi \text{ and}$$
$s = \left(\frac{gJ\mu_B}{k_BT}\right)\sin\theta\sin\varphi$. The magnetic moment per molecule is:

$$m = \frac{1}{Z}\left(\frac{4gJ\mu_B\sin\theta}{k_BT}\right)\left[(\sqrt{3}\sin\varphi\,\sinh(\sqrt{3}sB)\cosh(cB) + \cos(\varphi)(\sinh(2cB) + \sinh(cB)\cosh(\sqrt{3}sB))\right] \tag{A.1}$$

Expanding m as:
$$m(T,B) = \chi_1(T)B + \chi_3(T)B^3 + \chi_5(T)B^5 \tag{A.2}$$

and averaging over θ and φ, one obtains the angle averaged susceptibilities of eqs.(6-8). The averaging of $\chi_1(T)$ and $\chi_3(T)$ is simple since they do not depend on φ. The averaging of $\chi_5(T)$ however is facilitated by computer algebra.